\documentclass{article}
\usepackage[utf8]{inputenc}
\usepackage{hyperref}
\usepackage{amsmath,amssymb, enumerate}

\title{Interpreting the Recent Upper Limit on the Gravitational Wave Background from the Parkes Pulsar Timing Array}

\begin{document}

\author{The NANOGrav Collaboration }
\maketitle

Shannon et al.~(25 Sept, 2015) report important constraints from the
Parkes pulsar timing array (PPTA) on the gravitational-wave background
(GWB) from supermassive black-hole binaries (SMBHBs) using data from
four millisecond pulsars.  We wish to clarify two points regarding
their paper.

The reported non-detection of the GWB, widely publicized as a case of
``missing'' GWs, is based on simple power-law models (1--3).  These
models assume the ``last parsec problem'' is optimally solved, meaning
SMBHBs evolve, without stalling, to orbital frequencies where PTAs are
sensitive, and that SMBHB environments (stars, gas) and eccentric
orbits do not diminish the lowest frequency GWB signals.
These effects have been an active area of research for several
years (2--5) and current non-detections are unsurprising.  With more
millisecond pulsars, NANOGrav has placed new constraints on the shape
of the GWB spectrum, and thereby on SMBHB environments (6).

Shannon et al.~conclude that GWB detection will require additional
precisely-timed pulsars such as PSR J1909$-$3744, or higher-cadence
observations targeted at GW frequencies above 0.2~yr$^{-1}$.  Yet
recent studies show that the best GWB detection strategy, regardless
of GWB spectrum details, is to observe many pulsars over the longest
possible timespans to maximize sensitivity to the lowest GW
frequencies (7, 8). PTAs are most sensitive at frequencies near the
inverse of the total observing timespan, with sensitivity decreasing
at higher frequencies. No known mechanism can alter the GWB spectrum
to allow a PTA detection at higher frequencies.

The International Pulsar Timing Array (IPTA) aims to assemble the
timing data necessary for GW detection by pooling contributions from
the Parkes PTA, NANOGrav, and the European PTA. This approach, which
leverages very high quality data from the Arecibo and Green Bank
Telescopes, should enable a detection of the GWB within 10 years (8).

\begin{enumerate}
\item A. H. Jaffe, D. C. Backer, Gravitational Waves Probe the Coalescence Rate
of Massive Black Hole Binaries, Astrophys. J. 583, 616 (2003).
\item A. Sesana, Insights into the astrophysics of supermassive black hole binaries from pulsar timing observations. Classical Quant. Grav. 30, 224014 (2013), arXiv:1307.2600 [astro-ph.CO].
\item S. T. McWilliams, J. P. Ostriker, F. Pretorius, Gravitational waves and stalled satellites from massive galaxy mergers at $z <= 1$. Astrophys. J. 789, 156 (2014), arXiv:1211.5377 [astro-ph.CO].
\item E. A. Huerta, S. T. McWilliams, J. R. Gair, S. R. Taylor, Detection of eccentric supermassive black hole binaries with pulsar timing arrays: Signal-to-noise ratio calculations, Phys. Rev. D 92, 063010 (2015), arXiv:1504.00928 [gr-qc].
\item L. Sampson, N. J. Cornish, S. T. McWilliams, Constraining the solution to the last parsec problem with pulsar timing. Phys. Rev. D 91, 084055 (2015), arXiv:1503.02662 [gr-qc].
\item Z. Arzoumanian et al., The NANOGrav Nine-year Data Set: Limits on the Isotropic Stochastic Gravitational Wave Background, arXiv:1508.03024 [astro-ph.GA].
\item X. Siemens, J. Ellis, F. Jenet, J. D. Romano, The stochastic background: scaling laws and time to detection for pulsar timing arrays, Classical Quant. Grav. 30, 224015 (2013), arXiv:1305.3196 [astro-ph.IM].
\item S. R. Taylor, M. Vallisneri, J. A. Ellis, C. M. F. Mingarelli, T. J. W. Lazio, R. van Haasteren, Are we there yet? Time to detection of nanohertz gravitational waves based on pulsar-timing array limits, arXiv:1511.05564 [astro-ph.IM]
\end{enumerate}

\end{document}